\documentclass[prl,reprint,twocolumn,showpacs,superscriptaddress,floatfix,aps,10pt]{revtex4-2}

\usepackage{amsmath, amsthm, amssymb}
\usepackage{dcolumn, bm, hyperref}
\usepackage{graphicx, subfigure, verbatim}
\usepackage{txfonts}    
\usepackage{newtxtext}
\usepackage{xcolor}
\usepackage{soul}
\usepackage{subfiles}

\renewcommand{\vec}[1]{{\bm{\mathrm{#1}}}}

\newcommand{\bra}[1]{\langle{#1}|}
\newcommand{\ket}[1]{|{#1}\rangle}

\usepackage{xcolor}

\begin{document}

\title{Harnessing magnetic octupole Hall effect to induce torque in altermagnets}

\author{Seungyun Han$^\ddagger$}
\email{hanson@postech.ac.kr}
\affiliation{Department of Physics, Pohang University of Science and Technology, Pohang 37673, Korea}
\author{Daegeun Jo$^\ddagger$}
\affiliation{Department of Physics and Astronomy, Uppsala University, P.O. Box 516, SE-75120 Uppsala, Sweden}
\affiliation{Wallenberg Initiative Materials Science for Sustainability, Uppsala University, SE-75120 Uppsala, Sweden}
\author{Insu Baek$^\ddagger$}
\affiliation{Department of Physics, Pohang University of Science and Technology, Pohang 37673, Korea}
\author{Peter M. Oppeneer}
\affiliation{Department of Physics and Astronomy, Uppsala University, P.O. Box 516, SE-75120 Uppsala, Sweden}
\affiliation{Wallenberg Initiative Materials Science for Sustainability, Uppsala University, SE-75120 Uppsala, Sweden}
\author{Hyun-Woo Lee}
\email{hwl@postech.ac.kr}
\affiliation{Department of Physics, Pohang University of Science and Technology, Pohang 37673, Korea}

\begin{abstract}
$d$-wave altermagnets have magnetic octupoles as their order parameters [Phys. Rev. X $\mathbf{14}$, 011019 (2024)]. We theoretically show that magnetic octupoles injected from outside generate torque on the $d$-wave altermagnets. The injection can be achieved by the magnetic octupole Hall effect in an adjacent layer. We calculate the magnetic octupole Hall conductivity of the heavy metal Pt and find a sizable value comparable to its spin Hall conductivity. Our work generalizes the spin Hall phenomenology (generation by heavy metals and detection by torque in ferromagnets) to the magnetic octupole Hall phenomenology (generation by heavy metals and detection by torque in altermagnets), which can be utilized to electrically control magnetic configurations of altermagnets.
\end{abstract}

\maketitle

{\it Introduction.--} Analysis based on the spin space group has recently proposed a new class of magnetic materials dubbed altermagnet (AM)~\cite{hayami2019, naka2019, yuan2020, igor2021, yuan2021, yuan2022_1, yuan2022_2}. An AM has zero net magnetization, similar to antiferromagnet (AFM), but it exhibits a spin splitting of its energy bands~\cite{smejkal2022_1, smejkal2022_2,lee2024,han2024}, which does not originate from the relativistic spin-orbit coupling (SOC) and can be larger ($\sim$ 1 eV)~\cite{ahn2019, smejkal2020} than the SOC-induced spin splitting~\cite{pekar1964, rashba1960, bychkov1984}. This nonrelativistic splitting of the AM can be utilized to generate spin current~\cite{hernandez2021, bose2022}, spin-splitting torque~\cite{karube2022, bai2023, guo2024}, giant magnetoresistance~\cite{smejkal2022_3} and anomalous Hall effect~\cite{smejkal2020, shao2021}.

Recently, magnetic multipoles have emerged as primary quantities in AMs~\cite{bhowal2024}. Contrary to ferromagnets (FMs), where the magnetic dipoles are the order parameter, $d$-wave AMs have the magnetic octupoles (MOs) as their leading-order order parameter. The MOs distinguish the $d$-wave from conventional AFMs and determine the nonrelativistic spin-splitting~\cite{bhowal2024}. Additionally, a Landau theory analysis on $d$-wave AMs~\cite{mcclarty2024} pointed out that their MO $\mathbf{O}_{ij}$ couples linearly to their N\'eel vector $\mathbf{N}$, where the vector direction of $\mathbf{O}_{ij}$ denotes the spin direction and the indices $ij$ contain the angular distribution information of the MO [Eq.~\eqref{MOoperator}]. This coupling $\mathbf{N}\cdot \mathbf{O}_{ij}$ is analogous to the coupling $\mathbf{M}\cdot \mathbf{S}$ between the local magnetization $\mathbf{M}$ and the spin $\mathbf{S}$ in FMs. This analogy motivates the possibility that MOs injected into an AM~\cite{tahir2023} may induce torque on the AM, just as spins injected into an FM induce torque on the FM. In this Letter, we show that the injection of MO from outside indeed generates the torque on an AM. The torque contains a non-staggered component, implying that it can rotate the Néel order of the AM. This torque can survive in a configuration where the spin-Hall-induced torque vanishes identically. This torque is our first main result. As a method for injecting the MO into an AM layer, we propose the MO Hall effect (MOHE) in heavy metals (HMs). HMs such as Pt are strong sources of spin Hall current. We find that Pt also generates a large MO Hall current, which is comparable in magnitude to the large spin Hall current generated by Pt. This is our second main result.

{\it Magnetic octupole.--} Magnetic multipoles naturally emerge in the interaction energy between magnetization density $\bm{\mu}(\mathbf{r})$ and an inhomogeneous magnetic field $\mathbf{H}(\mathbf{r})$~\cite{spaldin2008, suzuki2010, spaldin2013, suzuki2018, shitade2019, urru2022, tahir2023}: $H_{\text{int}}=-\int \bm{\mu}(\mathbf{r})\cdot \mathbf{H}(\mathbf{r})\ d^3r$ can be expanded into $-\left(\int \bm{\mu}(\mathbf{r})\ d^3r\right) \cdot \mathbf{H}(0)-\left(\int r_i\mu_j(\mathbf{r})\ d^3r\right)\partial_iH_j(0) - \frac{1}{2} \left(\int r_ir_j \mu_k(\mathbf{r})\ d^3r\right)\partial_i\partial_j H_k(0)$. In centrosymmetric AM, both magnetic-dipolar term $\left(\propto \int \bm{\mu}(\mathbf{r})\ d^3r\right)$ and magnetoelectric multipolar term $\left(\propto \int r_i\mu_j(\mathbf{r})\ d^3r\right)$ vanish, but the MO, $\left(\propto \int r_ir_j\mu_{k}(\mathbf{r})\ d^3r\right)$, can possess a finite value. Thus, the MO is the first symmetry-allowed magnetic multipole that acts as the order parameter of AM. To facilitate the evaluation of MO, we adopt the atomic-site MO operators~\cite{jackeli2009, iwazaki2023, footnote1}
\begin{equation}\label{MOoperator}
    O_{ij}^q \equiv \frac{1}{\hbar^2}\{L_i,L_j\}S_q,
\end{equation}
where $L_i$ is the orbital angular momentum operator, $S_q$ is the spin angular momentum operator, and $\{A, B\}=AB+BA$. Here, $\{L_i, L_j\}/\hbar^2$ represents the quadrupole density of electrons constructed by the spherical tensors~\cite{Shiina1997} and the generalized Stevens operator approach, $\{L_i, L_j\}/\hbar^2 = Cr_ir_j/a_0^2$~\cite{kuramoto2008, kusunose2008, santini2009, hayami2024, supple} where $C$ is constant and $a_0$ is the Bohr radius. Note that $O^q_{ij}$ is defined to share the same dimension as $S_q$, which facilitates the magnitude comparison between the spin Hall effect (SHE) and the MOHE.


\begin{figure}[t]
\includegraphics[width=8cm]{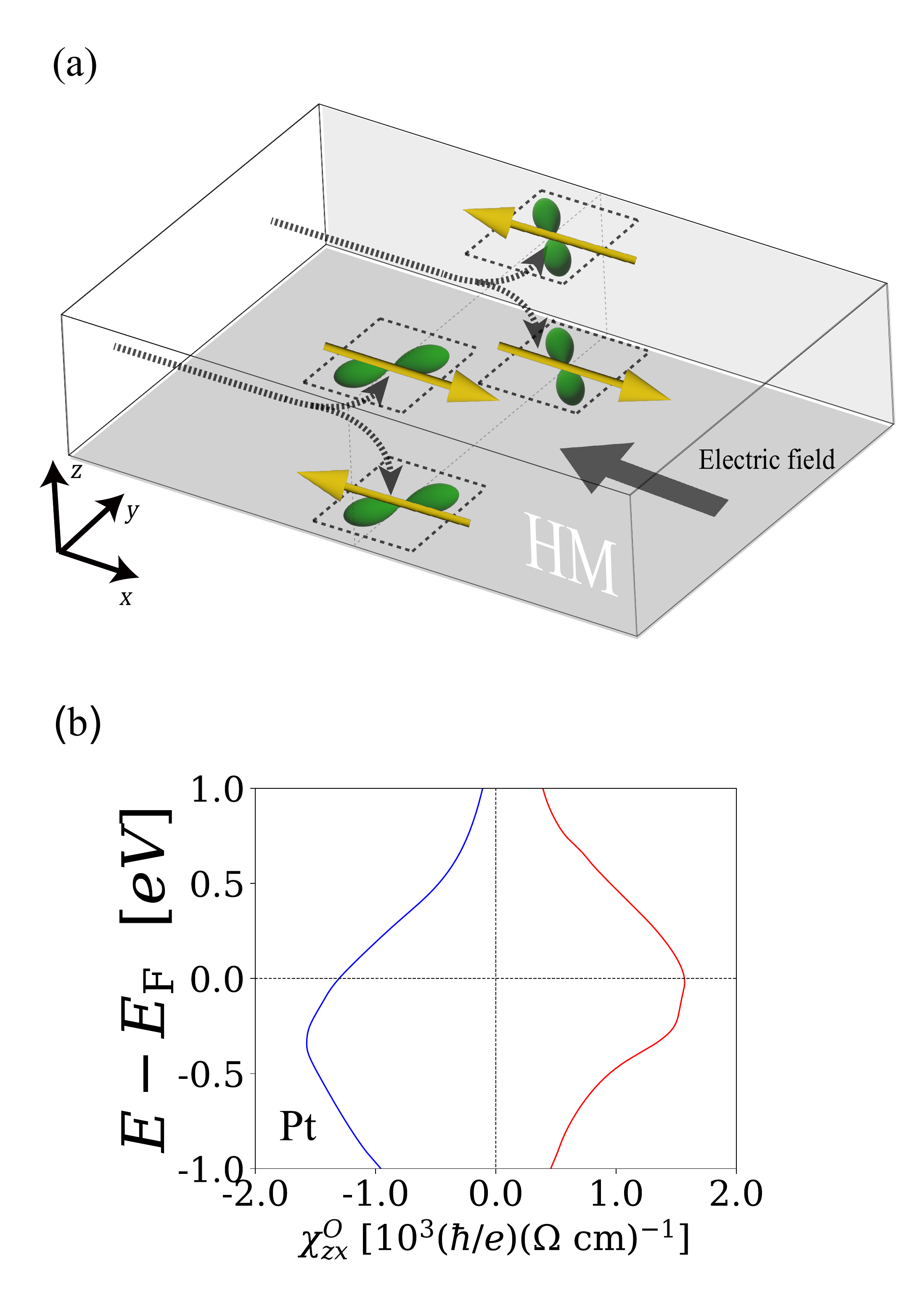}
 \caption{
(a) Illustration of MOHE. Under the external electric field along the $-x$ direction (black arrow), MOs with opposite signs flow along opposite $z$ directions, resulting in MOHE. The green color represents electron density, while the yellow arrows represent spin. Note that the depicted MO current is compatible with the mirror reflection symmetries with respect to both $zx$ and $zy$ planes. We use $p$-orbitals for illustration, but the same logic applies to $d$-orbitals, with electron density resembling that in the inset of Fig.~\ref{fig:3}.  (b) Computed magnetic octupole Hall conductivities $\chi_{zx}^{O_{xy}^x}$ (blue) and $\chi_{zx}^{O_{yz}^z}$ (red) for fcc Pt. 
 }
\label{fig:1}
\end{figure}

{\it Magnetic octupole Hall effect.--} We present our second main finding first: MOHE in HM. Within the linear response regime, the MO current density $J_j^{O^q_{mn}}$ carrying $O_{mn}^q$ is proportional to an electric field $\vec{\mathcal{E}}$,
\begin{equation}\label{MOHE}
    J_j^{O^q_{mn}}= \chi_{ji}^{O_{mn}^q} \mathcal{E}_i,
\end{equation}
where $J_j^{O^q_{mn}}$ is calculated by the MO current operator defined by $\{v_j, O_{mn}^q\}/2$, $\chi_{ji}^{O_{mn}^q}$ is the MO Hall conductivity (MOHC), and $v_j$ is the velocity operator. This is analogous to the SHE response, $J_j^{S_q} =\sigma_{ji}^{S_q} \mathcal{E}_i$, where $J_j^{S_q}$ represents the spin current density flowing in the $j$ direction with the spin along the $q$ direction. To be specific, we consider an electric field along the $x$-direction and compare the $z$-flows of MO currents and spin currents. For nonmagnetic materials with the mirror reflection symmetries with respect to the $zx$ and $zy$ planes, the SHE can generate only one component $J_z^{S_y}$, whereas the MOHE can generate multiple components, $J_z^{O_{xy}^x}$, $J_z^{O_{yz}^z}$, $J_z^{O_{xx}^y}$, $J_z^{O_{yy}^y}$, and $J_z^{O_{zz}^y}$ since MO and spin transform differently under the mirror reflections. Figure~\ref{fig:1}(a) illustrates $J_z^{O_{xy}^x}$. Note that the spin directions in $J_z^{O_{xy}^x}$ and $J_z^{O_{yz}^z}$ are perpendicular to the spin direction carried by the spin Hall current $J_z^{S_y}$. This will be useful in distinguishing the torque caused by MOHE from that caused by SHE~\cite{supple}. 
\begin{figure}[]
\includegraphics[width=8cm]{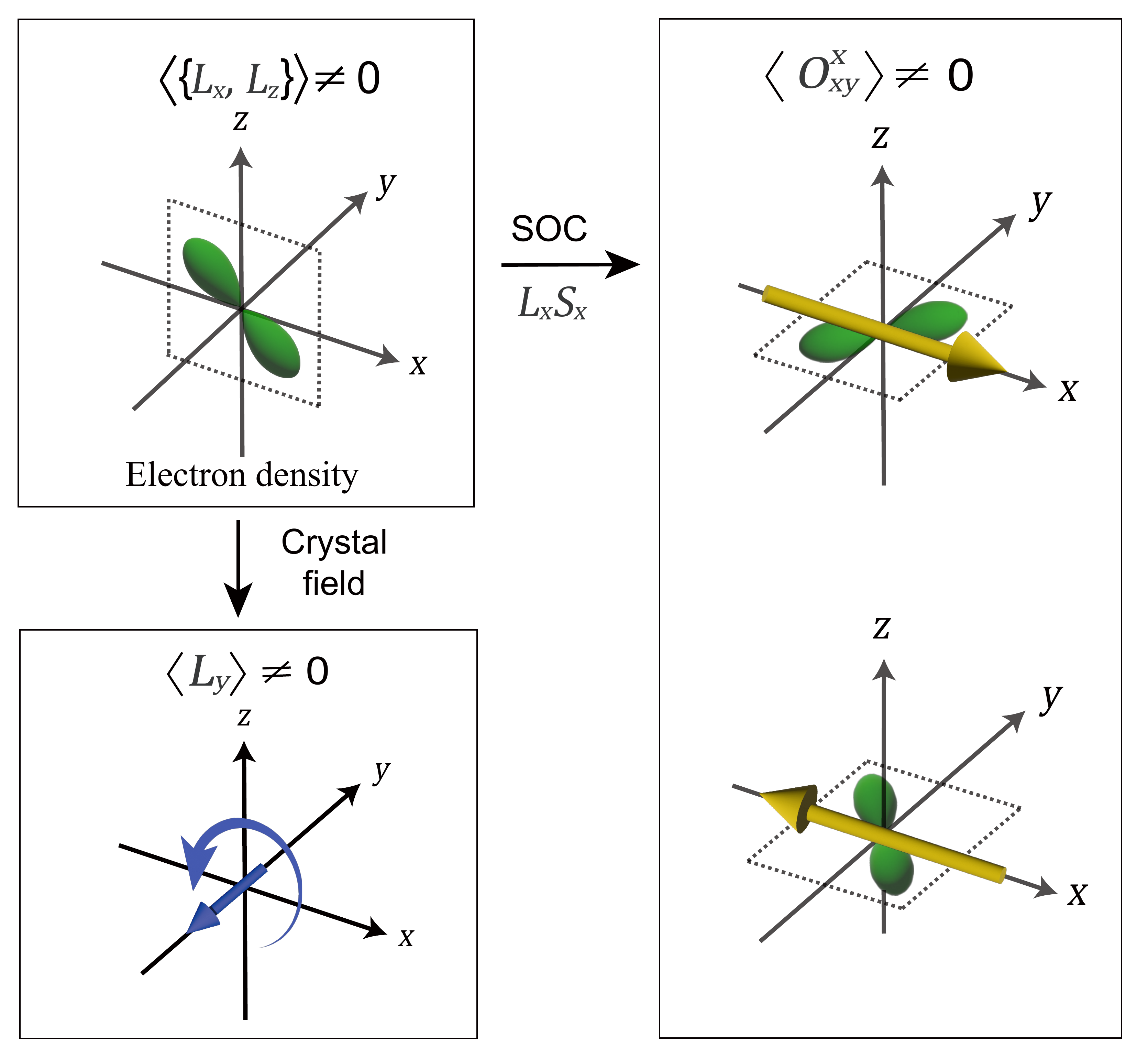}
 \caption{
The microscopic process of MO generation can be separated into two steps. First, due to the orbital texture, an electric field induces $\langle\{L_\alpha,L_\beta \}\rangle$ to deviate from its equilibrium expectation value (upper left panel). This is followed by the generation of additional nonequilibrium quantities: $\langle L_i\rangle$ (lower left panel) due to the crystal field [Eq.~\eqref{Lgeneartion}] and $\langle O_{ij}^q\rangle$ (right panel) due to the SOC [Eq.~\eqref{precession_a}]. When the two configurations in the right panel are combined, their net spin moment vanishes, but their net MO density is finite~\cite{supple}. Just as the generation of the nonequilibrium $\langle L_i\rangle$ leads to the OHE~\cite{han2022,go2018}, the generation of the nonequilibrium $\langle O_{ij}^q\rangle$ leads to the MOHE. Here, we illustrate only one particular index choice for each of $\langle\{L_\alpha, L_\beta \}\rangle$, $\langle L_i \rangle$, and $\langle O_{ij}^q\rangle$. In the right panel, we consider only $L_x S_x$ for SOC. Note that under this SOC, the nonequilibrium state with a positive $S_x$ rotates counterclockwise, while one with a negative $S_x$ rotates clockwise along the $x$ axis.
}
\label{fig:2}
\end{figure}


To confirm that MOHE exists in real materials, we perform the first-principles calculation for face-centered cubic (fcc) Pt~\cite{supple}. The MOHC $\chi^{O^q_{mn}}_{ji}$ is calculated through the Kubo formula within the linear response theory, which is given by
\begin{align}\label{MOkubo}
    \chi_{ji}^{O_{mn}^{q}} & = \frac{e}{\hbar} \sum\limits_{\mu \neq \nu} \int \frac{d^3 k}{(2 \pi)^3}  \\
&\quad\times (f_{\mu\textbf{k}}-f_{\nu\textbf{k}}) \hbar^2 \operatorname{Im}\Big[\frac{\bra{u_{\mu\textbf{k}}}{\frac{1}{2} \{v_j, O_{mn}^q\}}\ket{u_{\nu\textbf{k}}}\bra{u_{\nu\textbf{k}}}{v_i}\ket{u_{\mu\textbf{k}}}}{(E_{\mu\textbf{k}}-E_{\nu\textbf{k}})(E_{\mu\textbf{k}}-E_{\nu\textbf{k}} + i\Gamma)}\Big]\nonumber.
\end{align}
Here $\vert u_{\mu\mathbf{k}} \rangle$, $E_{\mu\mathbf{k}}$, and $f_{\mu\mathbf{k}}$ represent the periodic part of the Bloch eigenstate, the energy eigenvalue, and the Fermi-Dirac distribution function for the crystal momentum $\mathbf{k}$ and band index $\mu$, respectively, and $\Gamma=0.01$ eV is the energy level broadening. Figure~\ref{fig:1}(b) shows the allowed components $\chi_{zx}^{O_{xy}^x}$ (blue) and $\chi_{zx}^{O_{yz}^z}$ (red)~\cite{supple} of bulk Pt with respect to the Fermi energy $E$. The MOHC of fcc Pt can reach $\chi_{zx}^{O_{xy}^{x}} = -1303 \ (\hbar / e) (\Omega \ \rm{cm})^{-1}$ and $\chi_{zx}^{O_{yz}^{z}} = 1569 \ (\hbar / e) (\Omega \ \rm{cm})^{-1}$ at the true Fermi energy $E= E_{\text{F}}$. Note that these MOHCs share the same dimension as the spin Hall conductivity and are comparable to the spin Hall conductivity $\sigma_{zx}^{S_y} \sim 2000 \ (\hbar / e) (\Omega \ \rm{cm})^{-1}$ of Pt~\cite{guo2008}.

The mechanism of MOHE in time-reversal centrosymmetric systems resembles that of the orbital Hall effect (OHE)~\cite{jo2024,go2021}. To simplify the mechanism illustration, we consider a $p$-orbital Hamiltonian with finite SOC,
\begin{equation}\label{generalHamiltonian}
     \mathcal{H}(\mathbf{k})  = h_0 (\mathbf{k}) + \sum_{\alpha,\beta}h_{2,\alpha\beta}(\mathbf{k})\{L_\alpha,L_\beta\} + \lambda \mathbf{L}\cdot \mathbf{S},
\end{equation}
where $\lambda$ is the SOC strength. Here, $h_{2,\alpha \beta}(\mathbf{k})$ captures crystal field~\cite{han2022}, and its $\mathbf{k}$ dependence determines the orbital texture that is crucial for the generation of the OHE~\cite{go2018}. The mechanisms of both OHE and MOHE share the same first step: an electric field applied to systems with the orbital texture induces $\{L_{\alpha}, L_{\beta}\} $’s to deviate from their equilibrium expectation values (upper left panel in Fig.~\ref{fig:2}). The second steps of OHE and MOHE are different. For the OHE, the nonequilibrium $\{L_{\alpha}, L_{\beta}\}$’s generate the orbital angular momentum $L_i$, since 
\begin{equation}\label{Lgeneartion}
    \frac{dL_i}{dt}= \frac{[L_i,\mathcal{H}(\mathbf{k})]}{i\hbar}=A_{ijp}(\mathbf{k})\{L_j,L_p\},
\end{equation}
where $j,p$ are summed over, $A_{ijp}(\mathbf{k})$ is a combination of $h_{2,\alpha\beta}(\mathbf{k})$, and $\lambda$ = 0 is assumed for simplicity since the SOC is not necessary for the OHE. The second step of the OHE is illustrated in the lower left panel of Fig.~\ref{fig:2}. Now, we switch to the second step of the MOHE, where the nonequilibrium $\{L_{\alpha}, L_{\beta}\}$’s generate the MO, since

\begin{align}\label{precession_a}
    \frac{dO_{ij}^q}{dt}&=\frac{[O_{ij}^q,\mathcal{H}(\mathbf{k})]}{i\hbar} \\
&=\frac{\lambda}{4}\left(\epsilon_{jqm}\{L_i,L_m\}+\epsilon_{iqm}\{L_j,L_m\}\right) + \sum_{\mu,\nu} B^{\mu\nu}_{ijq}(\mathbf{k})L_\mu S_\nu, \nonumber 
\end{align}
where $\epsilon_{jqm}$ is the Levi-Civita symbol, $m$ is summed over, and $B^{\mu\nu}_{ijq}(\mathbf{k})$ is a combination of $\lambda$ and $h_{2,\alpha\beta}(\mathbf{k})$. The first term on the right-hand side of Eq.~\eqref{precession_a} is due to the SOC and amounts to the second step of the MOHE (the right panel of Fig.~\ref{fig:2}). We note that although the mechanism of the MOHE is illustrated for a $p$-orbital system, it is applicable to a wider class of systems, including $d$-orbital transition metals.
\begin{figure}[t]
	\includegraphics[scale=0.48]{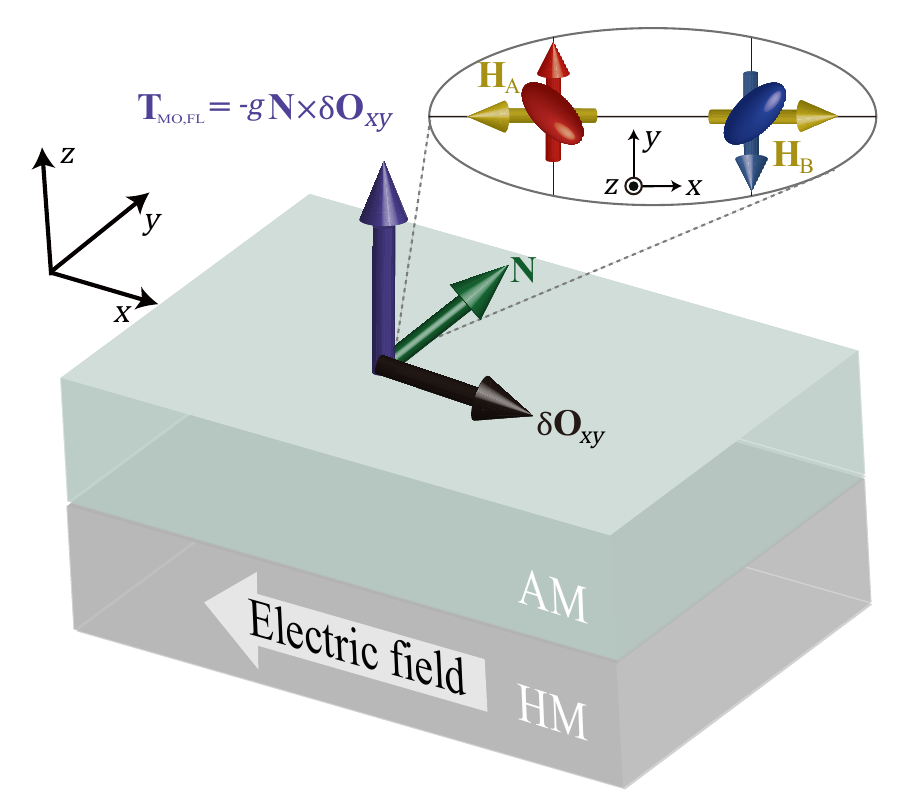}
	\caption{Illustration of field-like MOT mechanism. $\mathbf{N}$ is the N\'eel vector, $\delta \mathbf{O}_{xy}$ is the nonequilibrium MO density, and $\mathbf{T}_{\text{MO,FL}}$ is the field-like component of the resulting MOT. (Inset) The blue and red ellipses represent the spin density profiles of each sublattice in the AM, with blue and red arrows indicating their respective magnetization vectors. When $\delta\mathbf{O}_{xy}$ is induced in the AM via the MOHE in the HM, due to the correlation between electron density and spin moment of MO (right panel of Fig.~\ref{fig:2}), staggered effective fields $\mathbf{H}_{i}$ are induced, resulting in a non-staggered torque, $\mathbf{T}_{\text{MO,FL}}=-g \mathbf{N}\times\delta \mathbf{O}_{xy}$~\cite{supple}.}
	\label{fig:3}
\end{figure}

{\it Magnetic-octupole torque.--} We now discuss our first main finding: an MO current injected into an AM exerts torque on the AM. This MO torque (MOT) originates from the $\mathbf{N}\cdot\mathbf{O}_{ij}$ coupling in AM~\cite{mcclarty2024}. Since only a selected component $\mathbf{O}_{ij}$ of MO appears in the coupling, the injected MO current can generate the MOT only when the current carries the same indices $ij$. In CoF$_2$, MnF$_2$, and (001)-oriented RuO$_2$, what couples to $\mathbf{N}$ is $\mathbf{O}_{xy}$~\cite{smejkal2023, bhowal2024}, which shares the same indices ($i=x$, $j=y$) as $J_z^{O_{xy}^x}$ generated by the MOHE in Pt. Thus, the MOHE can generate the MOT on the AM in Pt/AM bilayers (Fig.~\ref{fig:3}).

Next, we infer the structure of the MOT by using the analogy to the conventional spin torque. When an extra spin $\delta\textbf{S}$ is injected into an FM with the $\mathbf{M} \cdot \mathbf{S}$ coupling, there arises the damping-like (DL) torque $\propto \mathbf{M} \times \left(\delta\mathbf{S} \times \mathbf{M}\right)$ and the field-like (FL) torque $\propto \mathbf{M} \times \delta\mathbf{S}$~\cite{haney2013}. This generalizes to the current-induced torque in HM/AFM bilayers, which can be again decomposed to the DL torque and the FL torque components. Specifically, the torque  $\vec{T}^\alpha$ on a magnetic atom $\alpha$ can be written as $\vec{T}^\alpha = T^\alpha_\mathrm{DL} \hat{\mathbf{m}}^\alpha  \times (\hat{\mathbf{p}} \times \hat{\mathbf{m}}^\alpha) + T^\alpha_\mathrm{FL} \hat{\mathbf{m}}^\alpha  \times \hat{\mathbf{p}}$~\cite{baltz2018antiferromagnetic}, where $\hat{\mathbf{m}}^\alpha$ is the unit vector of the magnetic moment at atom $\alpha$ and $\hat{\mathbf{p}}$ denotes the direction of $\delta \mathbf{S}$ (up to sign). Collinear AFMs with sublattices A and B can have both staggered and non-staggered torques. Here we focus on the non-staggered components (e.g., $T_\mathrm{DL}^\mathrm{A} =  T_\mathrm{DL}^\mathrm{B} $ and $T_\mathrm{FL}^\mathrm{A} =  - T_\mathrm{FL}^\mathrm{B}$) since they are more important for the magnetization dynamics in AFM. In conventional spin-orbit torque (SOT) geometry, where HM/AFM bilayers are stacked along the $\hat{\mathbf{z}}$ direction and an electric field is along $\hat{\mathbf{x}}$ direction, $\hat{\mathbf{p}}$ is along $\hat{\mathbf{y}}$~\cite{baltz2018antiferromagnetic}. Therefore, the non-staggered component $\vec{T}_{\mathrm{SO}}$ of $\mathbf{T}^\alpha$ can be expressed as 
	\begin{equation}\label{eq:t_so}
		\vec{T}_\mathrm{SO} = T_\mathrm{SO,DL} \hat{\mathbf{n}}  \times (\hat{\mathbf{y}} \times \hat{\mathbf{n}} ) + T_\mathrm{SO,FL} \hat{\mathbf{n}}   \times \hat{\mathbf{y}},
	\end{equation} 
where $\hat{\mathbf{n}}$ is the unit vector of the N\'eel vector, and $T_\mathrm{SO, DL}$ and $T_\mathrm{SO, FL}$ are the coefficients for DL and FL SOTs, respectively. 

We are now ready to discuss the current-induced torque in HM/AM bilayers, where additional contribution [Eq.~\eqref{eq:t_mo}] can arise in addition to $\mathbf{T}_{\mathrm{SO}}$ [Eq.~\eqref{eq:t_so}] since the HM layer can exhibit not only the SHE but also the MOHE. For instance, the HM layer with a cubic symmetry generates the MO Hall current carrying ${O_{xy}^x}$ [Fig.~\ref{fig:1}(b)]. Thus, for the HM/AM bilayer with the MO ordering $\mathbf{O}_{xy}$ in AM, the non-staggered component $\vec{T}_{\mathrm{MO}}$ of the MOT is expected to be as follows:
    \begin{equation}\label{eq:t_mo}
        \vec{T}_\mathrm{MO} = T_\mathrm{MO,DL} \hat{\mathbf{n}}  \times (\hat{\mathbf{x}} \times \hat{\mathbf{n}} ) + T_\mathrm{MO,FL} \hat{\mathbf{n}}   \times \hat{\mathbf{x}},
    \end{equation} 
where $T_\mathrm{MO, DL}$ and $T_\mathrm{MO, FL}$ are the coefficients for DL and FL MOTs, respectively. The FL MOT is illustrated in Fig.~\ref{fig:3}. Interestingly, $\mathbf{T}_{\mathrm{MO}}$ is finite even when $\hat{\mathbf{n}}  = \hat{\mathbf{y}}$, in stark contrast to $\mathbf{T}_{\mathrm{SO}}$ [Eq.~\eqref{eq:t_so}], which vanishes when $\hat{\mathbf{n}} \parallel \hat{\mathbf{y}}$. 

\begin{figure*}[t]
	\includegraphics[scale=0.5]{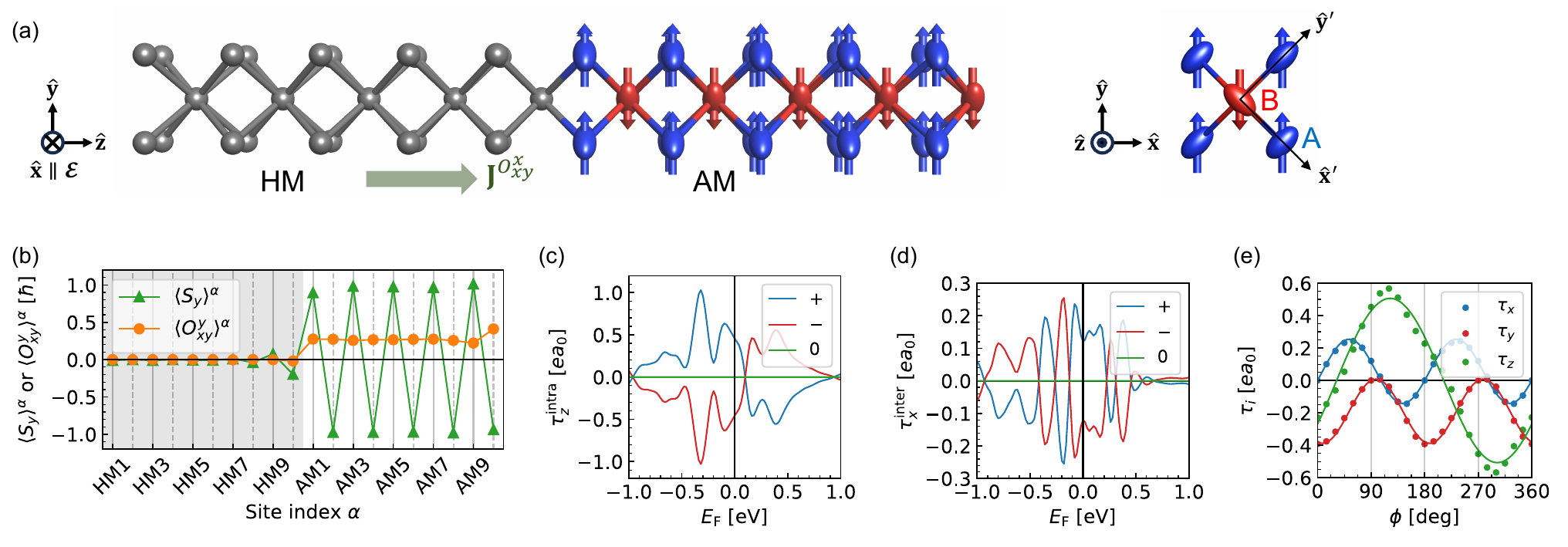}
	\caption{
		(a) Side (left panel) and top (right panel) views of an HM/AM bilayer. (b) The spin $\langle S_y \rangle^\alpha$ (green triangles) and atomic MO $\langle O_{xy}^y \rangle^\alpha$ (orange circles) expectation values in equilibrium, projected onto atomic sites $\alpha$. (c),(d) The torkances (c) $\tau_z^{\mathrm{intra}}$ and (d) $\tau_x^{\mathrm{inter}}$ as a function of the Fermi energy $E_\mathrm{F}$ for $\Delta_\mathrm{CF} = 1.0$ eV (blue), $-1.0$ eV (red), and 0 eV (green lines). (e) The $x$-, $y$-, and $z$-components (blue, red, and green dots, respectively) of the torkance as a function of the N\'eel vector angle $\phi$. Solid lines indicate the fitting by Eq.~\eqref{eq:tau_angle}.
	}
	\label{fig:4}
\end{figure*}

{\it Numerical calculation of magnetic octupole torque.--} To verify that the MOHE can induce the MOT, we perform numerical calculations of the current-induced torque in an HM/AM bilayer that consists of 10 atomic layers of HM and 10 atomic layers of AM with both materials in the body-centered cubic structure stacked along the [001] direction [left panel in Fig.~\ref{fig:4}(a)]. The $d$-wave AM layer is composed of two magnetic sublattices, denoted by A and B, with antiferromagnetic spin ordering, where their local crystal structures are related by the $C_{4z}$ rotational symmetry. Each atom can host three $t_{2g}$ orbitals $d_{x^\prime y^\prime}$, $d_{y^\prime z^\prime}$, and  $d_{z^\prime x^\prime}$, where the \textit{z}$^{\prime}$ direction is parallel to the \textit{z} direction but the \textit{x}$^{\prime}$ and \textit{y}$^{\prime}$ directions are 45$^{\circ}$ rotated from the \textit{x} and \textit{y} directions [right panel in Fig.~\ref{fig:4}(a)]. The tight-binding Hamiltonian $\mathcal{H}$ of the HM/AM bilayer is written as 
	\begin{align}\label{eq:tb_hamiltonian}
		\mathcal{H} & 
		=  \sum_{\langle \alpha,\beta \rangle , m,n, \sigma } t_{\alpha\beta,mn} c_{\alpha m\sigma}^\dagger c_{\beta n\sigma}
		+ \sum_{\alpha, n, \sigma } \mu_\alpha c_{\alpha n\sigma}^\dagger c_{\alpha n\sigma}   \\
		& \mathrel{\phantom{=}} 
		+  \frac{\lambda_\mathrm{SO}}{\hbar^2} \sum_{\substack{\alpha \in \mathrm{HM} \\ m,n, \sigma, \sigma' } }
		c_{\alpha m\sigma}^\dagger \mathbf{L}_{mn} \cdot \mathbf{S}_{\sigma \sigma'} \; c_{\alpha n\sigma'} \nonumber \\
          &\mathrel{\phantom{=}}\mathrel{\phantom{=}} 
		- \frac{J_\mathrm{sd}}{\hbar} \sum_{\substack{\alpha \in \mathrm{AM}  \\ n, \sigma, \sigma' }} \hat{ \mathbf{s} }_\alpha \cdot c_{\alpha n\sigma}^\dagger \mathbf{S}_{\sigma \sigma'} \; c_{\alpha n\sigma'}+   \sum_{\substack{\alpha \in \mathrm{AM} \\ n, \sigma }} 	\Delta_{\alpha n} \,	c_{\alpha n\sigma}^\dagger c_{\alpha n\sigma} \nonumber 
	\end{align}
where $c_{\alpha n\sigma}^\dagger$ ($c_{\alpha n\sigma}$) is the electron creation (annihilation) operator at site $\alpha$ with orbital index $n = x^{\prime}y^{\prime},y^{\prime}z^{\prime},z^{\prime}x^{\prime}$ and spin $\sigma = \uparrow,\downarrow$ along the \textit{z} direction. The first term of $\mathcal{H}$ describes the electron hopping, where $\langle \alpha,\beta \rangle$ indicates the nearest-neighbor sites $\alpha$ and $\beta$. The hopping parameter $t_{\alpha \beta,mn}$ is obtained by the Slater-Koster method~\cite{slater1954simplified}, with the parameter $t_\pi = 0.5$ eV describing $\pi$-bonding between $d$ orbitals. The second term corresponds to the site-dependent chemical potential: $\mu_\alpha = 0.2$ eV and $-0.2$ eV for AM and HM atoms, respectively. The third term describes the SOC of HM atoms with the SOC strength $\lambda_\mathrm{SO} = 0.1$ eV. The orbital angular momentum operator $\mathbf{L}$ is locally defined with the $d$ orbitals as bases. The fourth term represents the exchange interaction in AM between the local spin ${\hat{\mathbf{s}}}_{\alpha}$ and the conduction electron spin $\mathbf{S}$ in the mean-field approximation. The exchange energy $J_\mathrm{sd}$ is 1.0 eV and $\hat{\mathbf{s}}_\alpha = \eta_\alpha \hat{\mathbf{y}}$ is the unit vector of the local spin with the sublattice index $\eta_\alpha = +1(-1)$ for the AM atom $\alpha$ in the sublattice A(B). The last term describes the sublattice-dependent orbital splitting in AM, giving rise to altermagnetism. Here we consider a distorted octahedral crystal field splitting:  $\Delta_{\alpha,x^\prime y^\prime} = - \vert \Delta_\mathrm{CF} /2 \vert$ and $\Delta_{\alpha,y^\prime z^\prime} = - \Delta_{\alpha,z^\prime x^\prime} = \eta_\alpha \Delta_\mathrm{CF} /2 $ with $\Delta_\mathrm{CF} = 1.0$ eV.
This term sets $\mathbf{O}_{xy}$ as the order parameter that characterizes the AM [right panel in Fig.~\ref{fig:4}(a)]. Note that the coupling $\mathbf{N}\cdot\mathbf{O}_{xy}$ does not appear explicitly in $\mathcal{H}$ [Eq.~\eqref{eq:tb_hamiltonian}]. Instead, it emerges from the combined action of the last term ($\Delta_{\rm{CF}}$) and the fourth term ($J_{sd}$), which can be verified from the projection of $\mathcal{H}$ to spin-split bands~\cite{supple} or from the alignment of the expectation value $\langle {\mathbf{O}}_{xy} \rangle$ to $\hat{\mathbf{n}} \sim \eta_\alpha \langle \hat{\mathbf{s}}_\alpha \rangle=\hat{\mathbf{y}}$ [Fig.~\ref{fig:4}(b)]~\cite{bhowal2024, mcclarty2024}.

Employing the linear response theory, we calculate the torkance $\vec{\tau} = \mathbf{T}/ \mathcal{E} $, namely the torque per electric field $\vec{\mathcal{E}} = \mathcal{E} \hat{\mathbf{x}}$. The torkance can be decomposed into intraband and interband contributions (cf.~\cite{freimuth2014}), $\tau_{i} = \tau_{i}^\mathrm{intra} + \tau_{i}^\mathrm{inter}$ ($i=x,y,z$):
	\begin{subequations}\label{t_total}
	\begin{align}
		\tau_{i}^\mathrm{intra} = & \frac{e\hbar}{\Gamma} \int d^2\mathbf{k} \sum_{\mu} \frac{\partial f_{\mu\mathbf{k}}}{ \partial E_{\mu\mathbf{k}}}
		\; \langle u_{\mu\mathbf{k}} \vert \mathcal{T}_i \vert u_ {\mu\mathbf{k}}\rangle \langle u_{\mu\mathbf{k}} \vert v_a \vert u_{\mu\mathbf{k}} \rangle, \label{t_intra} \\
		\tau_{i}^\mathrm{inter} = & e\hbar \int d^2\mathbf{k} \sum_{\mu \neq \nu} 
		(f_{\mu\mathbf{k}} - f_{\nu\mathbf{k}}) \nonumber \\
  & \times \mathrm{Im} \left[ \frac{\langle u_{\mu\mathbf{k}} \vert \mathcal{T}_i \vert u_ {\nu\mathbf{k}}\rangle \langle u_{\nu\mathbf{k}} \vert v_a \vert u_{\mu\mathbf{k}} \rangle }{(E_{\mu\mathbf{k}} - E_{\nu\mathbf{k}})(E_{\mu\mathbf{k}} - E_{\nu\mathbf{k}} + i\Gamma )} \right] \label{t_inter},
	\end{align}
	\end{subequations}
where $\mathcal{T}_i = (1/i\hbar)[S_i, \mathcal{H}(\mathbf{k})]$ is the torque operator exerted on the $i$-component spin, and the energy level broadening $\Gamma$ is assumed to be 25 meV. Here, $\mathcal{H}(\textbf{k})$ is the two-dimensional Fourier transform of $\mathcal{H}$ in Eq.~\eqref{eq:tb_hamiltonian}. The blue curves in Figs.~\ref{fig:4}(c) and \ref{fig:4}(d) represent the only nonvanishing torkance component $\tau_{z}^\mathrm{intra}$ and $\tau_{x}^\mathrm{inter}$. Since $\hat{\mathbf{n}}=\hat{\mathbf{y}}$, $\vec{T}_{\rm{SO}}$ [Eq.~\eqref{eq:t_so}] cannot contribute to the nonvanishing torkances, which we attribute to $\vec{T}_{\rm{MO}}$ [Eq.~\eqref{eq:t_mo}] with $\tau_{x}^\mathrm{inter} = T_\mathrm{MO,DL}/\mathcal{E}$ and $ \tau_{z}^\mathrm{intra} = - T_\mathrm{MO,FL}/\mathcal{E}$. To verify the MO-origin of this torque further, we repeat the calculation with a reversed sign of $\Delta_\mathrm{CF}$ in $\mathcal{H}$, which should reverse $\textbf{O}_{xy}$ and the torkances, while preserving $\hat{\mathbf{n}}$. The red curves in Figs.~\ref{fig:4}(c) and \ref{fig:4}(d) demonstrate that the torques’ signs are indeed reversed. On the other hand, if $\Delta_\mathrm{CF} = 0$ (AM reducing to a conventional AFM), all the torques disappear [green lines in Figs.~\ref{fig:4}(c) and \ref{fig:4}(d)]. These results validate that the calculated torkances for $\hat{\mathbf{n}}=\hat{\mathbf{y}}$ come entirely from the MOT. Next, we investigate how the MOT depends on the direction of $\hat{\mathbf{n}}$. For an arbitrary direction of $\hat{\mathbf{n}}$, both the SOT and MOT can occur. With $\hat{\mathbf{n}} = \cos\phi \, \hat{\mathbf{x}} + \sin\phi \, \hat{\mathbf{y}}$, Eqs.~\eqref{eq:t_so} and \eqref{eq:t_mo} predict the three components of the total torkance to exhibit the following angular dependencies:
	\begin{align}\label{eq:tau_angle}
	\tau_x & = - \tau_\mathrm{SO,DL} \sin \phi \cos \phi + \tau_\mathrm{MO,DL} \sin^2 \phi,  \nonumber \\
	\tau_y & = \tau_\mathrm{SO,DL} \cos^2 \phi - \tau_\mathrm{MO,DL} \sin \phi \cos \phi, \nonumber \\
	\tau_z & = \tau_\mathrm{SO,FL} \cos \phi - \tau_\mathrm{MO,FL} \sin \phi. 
	\end{align}
The numerical calculation results [Fig.~\ref{fig:4}(e)] from Eq.~\eqref{t_total} are in excellent agreement with this prediction with coefficient $\tau_\mathrm{SO,DL} = -0.38$, $\tau_\mathrm{SO,FL} = -0.27$, $\tau_\mathrm{MO,DL} = 0.11$, and $\tau_\mathrm{MO,FL} = -0.43$ in units of $ea_0$. This result indicates that the MOHE in the HM can generate a large MOT on the AM, which can be comparable to the conventional SOT.

{\it Discussion and outlook.--} A recent experiment~\cite{han2024} reported electrical 180$^{\circ}$ switching of N\'eel vector in a Pt/Mn$_5$Si$_3$ bilayer. Although the experiment is interpreted in terms of the SOT, we argue that the MOT is also relevant. So far, we have focused on an AM with the MO order $\mathbf{O}_{xy}$, for which the MO Hall current $J_z^{O_{xy}^x}$ generates the MOT on the AM. For an AM with a different crystal field structure, different MO orders may be relevant, for which different components of the MO Hall current can generate the MOT. Thus, the MOT depends on the crystal field structure of AM, which is a unique feature of AM. Additionally, our work can be generalized to $g$- and $i$-wave AMs. These AMs can be characterized by higher-order multipoles, and the injection of the multipoles from a neighboring layer can generate torque on the AMs. This implies that AM is a promising platform for multipoletronics~\cite{tahir2023}. Recently, $g$-wave AMs $\alpha$-Fe$_2$O$_3$, CrSb, and MnTe were reported~\cite{mcclarty2024, verbeek2024}. A search for materials with the higher-order multipole Hall effect is needed. Finally, we suggest that AM may be an interesting platform for orbitronics~\cite{jo2024, go2021}. Existing orbitronics studies examined similarities between the orbital angular momentum $\mathbf{L}$ and the spin $\mathbf{S}$, as well as their correlation. On the other hand, our work on AM deals with the correlation between $\{L_i, L_j\}$ and $\mathbf{S}$, which may have interesting implications for orbitronics since $\mathbf{L}$ and $\{L_i,L_j\}$ degrees of freedom are strongly interwined~\cite{han2022}.

\begin{acknowledgments}
We thank Dongwook Go, Suik Cheon, Sang-Wook Cheong, and Kyoung-Whan Kim for fruitful discussions. S.H., I.B., and H.-W.L. were financially supported by the National Research Foundation of Korea (NRF) grant funded by the Korean government (MSIT) (No. RS-2024-00356270). S.H. was financially supported by the NRF grant funded by the Korean government (No. RS-2024-00334933). D.J. and P.M.O. were supported by the Wallenberg Initiatie Materials Science for Sustainability (WISE) funded by the Knut and Allice Wallenberg Foundation. D.J. and P.M.O. further acknowledge support from the K. and A. Wallenberg Foundation (Grants No. 2022.0079 and 2023.0336) and from the National Academic Infrastructure for Supercomputing in Sweden (NAISS) at NSC Link\"oping partially funded by the Swedish Research Council through grant agreement No. 2022-06725. Supercomputing resources, including technical support, were provided by the Supercomputing Center, Korea Institute of Science and Technology Information (Contract No. KSC-2023-CRE-0390). S. Han, D. Jo, and I. Baek contributed equally to this work. 
\end{acknowledgments}

\end{document}